\documentstyle[preprint,prl,aps]{revtex}
\input psfig
\newcommand{\LC}{$\Lambda_c^+ $}
\newcommand{\SP}{$\Sigma^+ $}
\newcommand{\LAMPI}{$\Lambda\pi^+ $}
\newcommand{\SIGPI}{$\Sigma^+\pi^0 $}
\newcommand{\RA}{\rightarrow}
\newcommand{\COST}{$\cos\Theta $}
\newcommand{\aLC}{\alpha_{\Lambda_c}}
\newcommand{\A}{$A$\ }
\newcommand{\B}{$B$\ }
\newcommand{\LClampi}{$\Lambda_c^+ \RA \Lambda\pi^+ $\ }
\newcommand{\LCsigpi}{$\Lambda_c^+ \RA \Sigma^+\pi^0 $\ }

\begin{document}
\tighten
\draft

\preprint{\parbox[b]{1.2in}{CLNS 95/1319 \\ CLEO 95-1 } }

\title{ Measurement of the Decay Asymmetry Parameters in
$\Lambda_c^+ \rightarrow \Lambda\pi^+$ and
$\Lambda_c^+ \rightarrow \Sigma^+\pi^0 $ }

\author{
M. Bishai,$^{1}$ J.~Fast,$^{1}$ E.~Gerndt,$^{1}$ J.W.~Hinson,$^{1}$
R.L.~McIlwain,$^{1}$ T.~Miao,$^{1}$ D.H.~Miller,$^{1}$
M.~Modesitt,$^{1}$ D.~Payne,$^{1}$ E.I.~Shibata,$^{1}$
% ...
}

\address{
\bigskip 
{\rm (CLEO Collaboration)}\\  
\newpage 
$^{1}${Purdue University, West Lafayette, Indiana 47907}\\
% ...
\bigskip 
}        

\date{February 8, 1995 }

\maketitle

\begin{abstract}

We have measured the weak decay asymmetry parameters ($\aLC $) for two
\LC\   decay modes.  Our measurements are
$\aLC = -0.94^{+0.21+0.12}_{-0.06-0.06} $ for the decay mode
$\Lambda_c^+ \rightarrow \Lambda\pi^+ $ and
$\aLC = -0.45\pm 0.31 \pm 0.06$ for the decay mode $\Lambda_c \rightarrow
\Sigma^+\pi^0 $. By combining these measurements with the previously measured
decay rates, we have extracted the parity-violating and parity-conserving
amplitudes. These amplitudes are used to test models of nonleptonic charmed
baryon decay.

\end{abstract}


The weak decays of charmed baryons are more complex than those of charm
mesons, since non-spectator effects like W-exchange and internal
W-emission are significant.  The measured differences in the lifetimes
of $\Lambda_c^+ $, $\Xi^+_c $, and $\Xi_c^0 $\cite{PDG} and
measurements of branching ratios in various exclusive decay modes such
as $\Lambda_c^+ \rightarrow \Xi^0 K^+ $\cite{Yelton} and $\Lambda_c^+
\rightarrow \Sigma^+\pi^0 $\cite{Procario} demonstrate this.
Currently, the strong interaction effects in these decays can only be
calculated using models, so it is critical to provide as much
experimental guidance as possible. Unlike many of the observed charmed meson
decays, the degree of parity violation in $\Lambda_c^+ $ decays to
a pseudoscalar and hyperon is an additional experimental observable.

For the decays we are studying, \LClampi and \LCsigpi \cite{Charge},
the parity violation is manifested by a polarization of the hyperon
($\Lambda $ or $\Sigma^+ $).  Because the hyperon decay also violates
parity, the hyperon polarization (helicity) can be
measured by its decay angular distribution.

\begin{figure}
\centerline{
   \psfig{bbllx=120pt,bblly=510pt,bburx=470pt,bbury=700pt,file=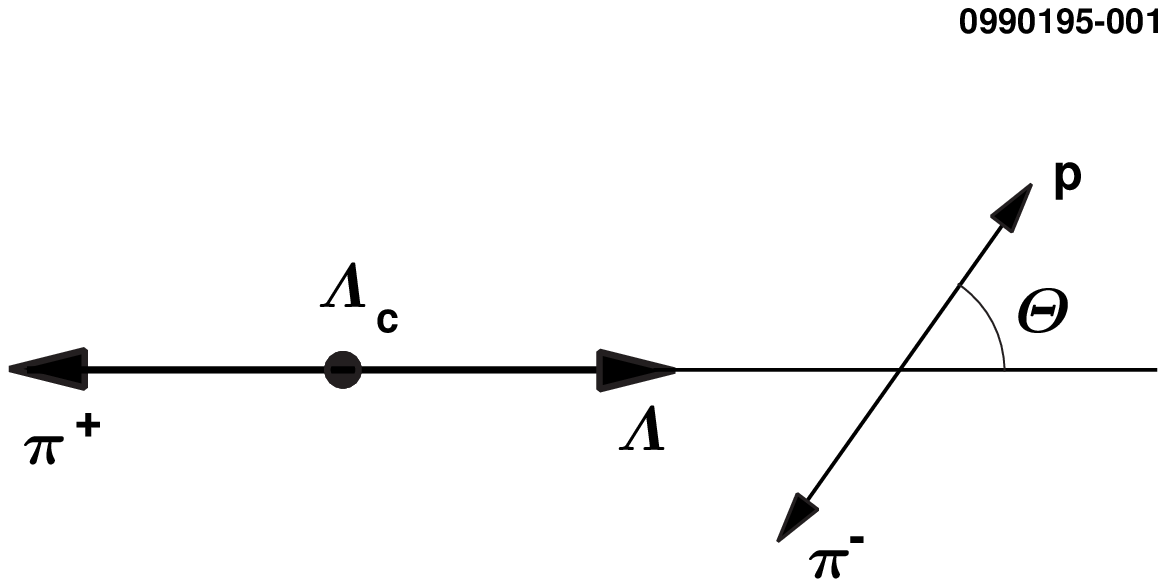}
}
  \caption{ Definition of the angle $\Theta$ in \LAMPI\   decay }
  \label{fig:helicity}
\end{figure}

We measure the helicity angle $\Theta $ in the hyperon's rest frame
between the proton's momentum and the direction opposite the \LC\
momentum, as shown in Figure~\ref{fig:helicity}.  The distribution for
this angle is given by:

\begin{equation}
 R = \frac{1}{2} (1 + \aLC \alpha \cos{\Theta} ),
 \label{eqn:Rlambda}
\end{equation}
where $\alpha $ and $\aLC $, the weak decay asymmetry parameters of the hyperon
and of the \LC, are defined as:
\begin{equation}
 \alpha  =  2 \frac{\kappa{\rm Re}(AB^{*})}{|A|^{2}+\kappa^{2}|B|^{2}},
\end{equation}
where $\kappa = p_f/(E_f+m_f) $ with $p_f $, $E_f $ and $m_f $ being
the momentum, energy and mass of the final baryon.  Here \A is the
parity-violating amplitude, and \B is the parity-conserving amplitude
in that particular decay.  The hyperon asymmetry parameter $\alpha $
is known, and $\aLC $ will be our measurement.

The data were collected with the CLEO II detector at the Cornell
Electron Storage Ring (CESR), which ran at and below the $\Upsilon
(4S) $ resonance.  The CLEO II detector is a solenoidal-magnet
spectrometer and electromagnetic calorimeter. The central drift
chamber measures a charged particle's momentum and its specific
ionization, which is used for particle identification. The
time-of-flight system provides additional particle identification
information. The calorimeter consists of 7800 CsI(Tl) crystals located
inside the magnet.  It has high efficiency, fine segmentation, and
excellent energy resolution, allowing us to reconstruct $\Sigma^+ $
hyperons through their decay to $p\pi^0 $.  A complete description of
the detector can be found elsewhere \cite{NIM}.  We employ a
GEANT\cite{GEANT} based detector simulation for our Monte Carlo (MC).

The total integrated luminosity for the data sample is 1.9 fb$^{-1}$,
corresponding to about two million $e^{+} e^{-} \RA c \bar{c}$ events.
The hadronic event selection requires at least three charged tracks,
visible energy greater than 0.15 of the energy in the center of mass, and a
distance less
than 5.0 cm along the beam direction between the reconstructed primary
vertex and the interaction point.

Parity violation in the \LClampi decay mode has been observed before
\cite{CLEOasy,ARGUSasy}. The $\Lambda $ candidates are reconstructed
from their $p\pi^- $ decay mode which has a branching fraction of
64.1\%. We search for a pair of oppositely charged tracks which
intersect at a radial distance of greater than 1 mm from the primary
vertex. The higher momentum particle is assumed to be the proton, and
the specific ionization (dE/dx) measurement is required to be
consistent with that hypothesis.  A $\chi^2 $ combining the distance
between the tracks in $z$ at the intersection point and the
extrapolated impact parameter of the $\Lambda $ candidate with the primary
vertex is required to be consistent with that of a $\Lambda $ coming from
the primary vertex.

Candidates within 5 MeV/c$^2$ of the known $\Lambda $ mass are
combined with other positively charged tracks in the event.
Combinatoric backgrounds are reduced by requiring that \LAMPI\
combinations have $x_p > 0.5 $, where $x_p = p/p_{max}$ and $p_{max} =
\sqrt{E_{beam}^2 - m_{\Lambda_c^+}^2}$, and that the $\pi^+ $ must be
within $90^\circ $ of the \LC\ candidate's direction.  The
distribution is fitted to a $3^{\rm rd}$ order Chebyshev polynomial
and a Gaussian, fixed to the MC width of 7.9 MeV/c$^2$, plus a
box-shaped function obtained from the MC to model the reflection from
the decay $\Lambda_c^+ \RA \Sigma^0\pi^+ $, where $\Sigma^0
\rightarrow \Lambda\gamma $ and the $\gamma $ is ignored.  We observe
$414 \pm 30$ events.

\begin{figure}
 \centerline{
\psfig{bbllx=60pt,bblly=330pt,bburx=550pt,bbury=700pt,file=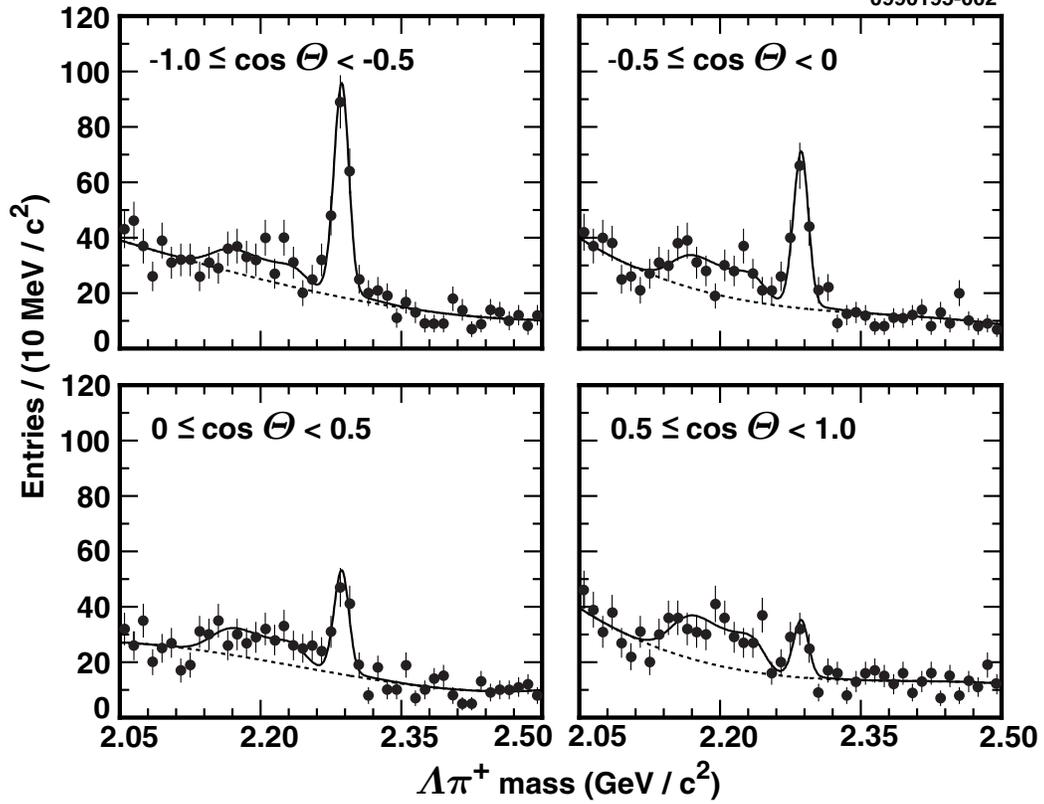,width=5.5in}
 }
  \caption{ $\Lambda\pi^+ $ Invariant mass distribution in four $\cos\Theta$
bins}
  \label{fig:lampicos}
\end{figure}

We divide the data into four $\cos\Theta$ bins and fit the \LAMPI\
invariant mass distributions with the mean and the width fixed to that
of the overall fit.  Since the asymmetry of $\Lambda_c^+ \RA
\Sigma^0\pi^+ $ can be different from that of \LClampi the relative
size of the $\Lambda_c^+ \RA \Sigma^0\pi^+ $ reflection to the
\LClampi signal is not fixed.  The invariant mass distributions can be
seen in Figure~\ref{fig:lampicos}. A clear decrease in the number of
$\Lambda\pi^+ $ events occurs as \COST\ increases.

\begin{figure}
 \centerline{
\psfig{bbllx=90pt,bblly=390pt,bburx=490pt,bbury=720pt,file=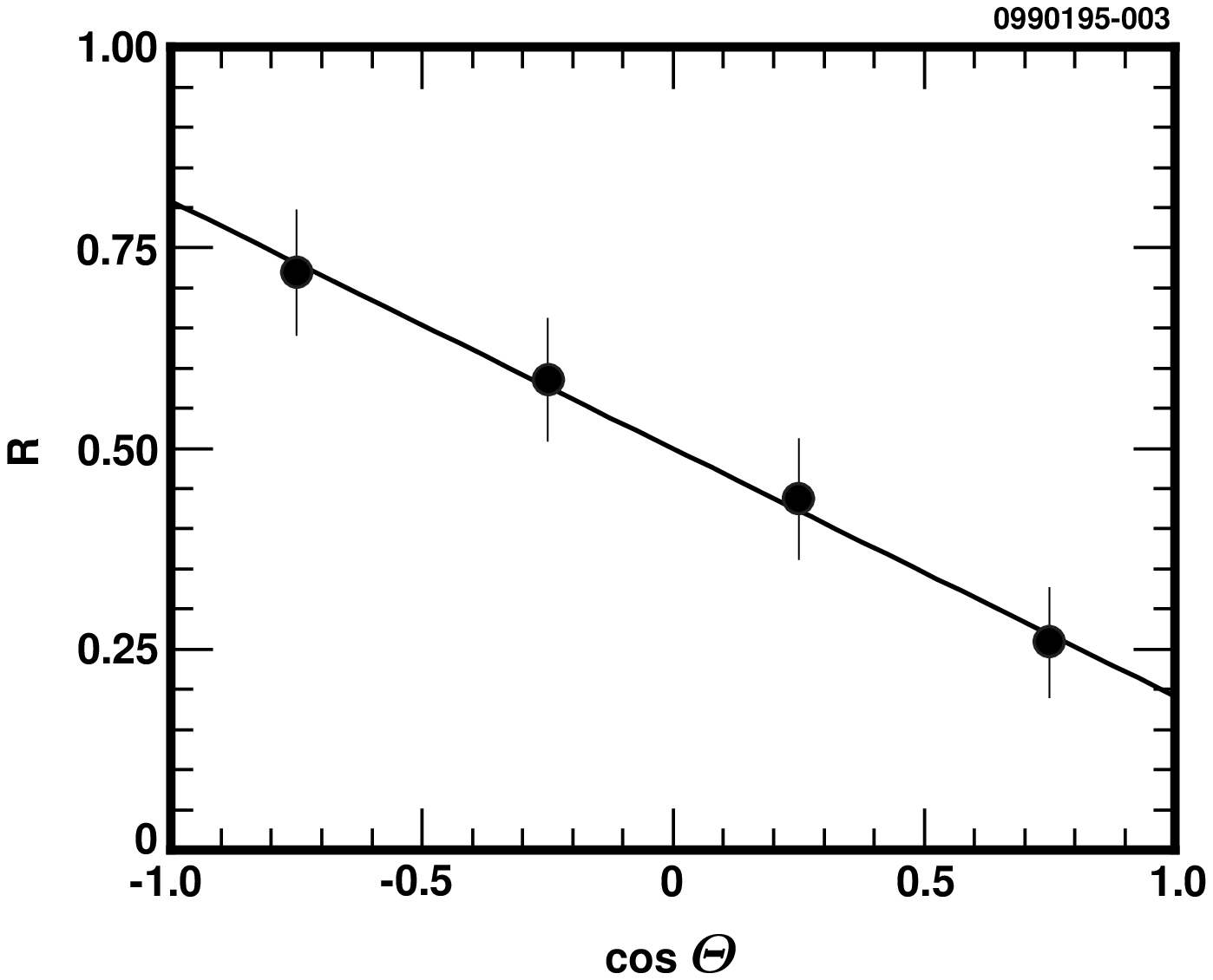,width=5.0in}
 }
 \caption{ Efficiency corrected yields vs $\cos\Theta$ for decay \LClampi}
 \label{fig:lampiasy_show}
\end{figure}

We calculate the efficiency for each \COST\ bin.  The yields are then
efficiency corrected, normalized, and plotted as a function of \COST,
as shown in Figure~\ref{fig:lampiasy_show}.  The slope is $ -0.30 \pm
0.07$, with a confidence level of $98\%$. Knowing that $
\alpha_\Lambda = 0.64 \pm 0.01$, we obtain $\alpha_{\Lambda_c} =
-0.94\ ^{+\ 0.21}_{-\ 0.06} $. The errors are statistical only, and
they are asymmetric because $\aLC $ cannot be smaller than $-1.0$.

We have estimated a variety of systematic errors. The first is due to
imperfect modeling of the efficiency.  The MC predicted efficiency as a
function of \COST\ varies by a small amount, due to the loss of efficiency for
low momentum pions from $\Lambda $ decay, which curl up in the magnetic
field. We have used pions from $D^{*+} $ decays to study the accuracy of our
modeling of the efficiency for low momentum pions.  We also have varied the
assumed fragmentation function of the \LC's in our MC, since this changes the
momentum distribution of the pions from $\Lambda $ decay. We have varied the
fitting procedure for finding the $\Lambda\pi^+ $ yield by varying the
background shape and by either modeling the $\Sigma^0\pi^+ $ bump or excluding
the region.  Combining all sources of systematic error in quadrature, we find
a total systematic error on $\aLC $ of $\pm 0.12$ dominated by the fit to the
$\Lambda\pi^+ $ invariant mass distributions.
Our result is consistent with previous measurements by CLEO\cite{CLEOasy}
and ARGUS\cite{ARGUSasy} of $-1.0^{+0.4}_{-0.0} $ and $-0.96\pm 0.42$
respectively.

We search for $\Sigma^+ $ candidates in the $p\pi^0$ decay mode where the
$\pi^0 $ subsequently decays to $\gamma\gamma $. A more detailed description
of our \SP\ reconstruction technique can be found elsewhere \cite{Procario}.
Proton candidates are identified by their ionization loss in the central drift
chamber and/or their time-of-flight.  The $\Sigma^+ $ is relatively long lived
($c\tau = 2.40$ cm), and decays a measurable distance from the primary
interaction vertex.  Hence protons from the $\Sigma^+$ decay have large impact
parameters with respect to the primary vertex.  We require that the proton
impact parameter in the ($r,\phi$) plane be greater than 0.6 mm.

The $\pi^0 $ candidates are formed by pairing energy clusters in the
calorimeter that are not matched to charged tracks, have an energy of at least
30 MeV, and have at least one of the clusters in the highest resolution
portion of the calorimeter ($|\cos \theta | < 0.71$), where $\theta $ is the
angle with respect to the beam axis.  The photon momenta are
then adjusted by a kinematic fit constrained by the known $\pi^0 $ mass and by
the assumption that the photons originate at the primary vertex. Candidates
with momentum greater than 100 MeV/c are kept.

The $\Sigma^+ $ candidates are identified with an iterative method that finds
an estimated decay point and calculates the $\Sigma^+ $ four-momentum assuming
the decay occurred there.  The $\pi^0 $'s are refit assuming the photons
originate at this decay vertex. This procedure improves our $\Sigma^+$ mass
resolution and signal-to-background ratio.  We choose those $p\pi^0 $
combinations within 15 MeV/c$^2$ of the nominal $\Sigma^+ $ mass to be our
$\Sigma^+$ candidates.

These \SP\ candidates are combined with $\pi^0 $'s. We require the
$\Sigma^+\pi^0 $ candidates to have $x_p >
0.5 $ and that the angle between the $\pi^0$ and the \LC\
momentum vectors in the lab frame be less than $90^\circ$. This is
identical to the procedure  followed for the \LAMPI\ case. We also require the
$\pi^0 $s from the \LC\ to have mo4mentum above 500 MeV/c to further
suppress combinatoric backgrounds. The distribution is fitted to a
$3^{\rm rd}$ order Chebyshev polynomial and a Gaussian, fixed to the
MC width of 20.8 MeV/c$^2$.  We observe $89 \pm 14$ events.

\begin{figure}
 \centerline{
\psfig{bbllx=60pt,bblly=330pt,bburx=550pt,bbury=700pt,file=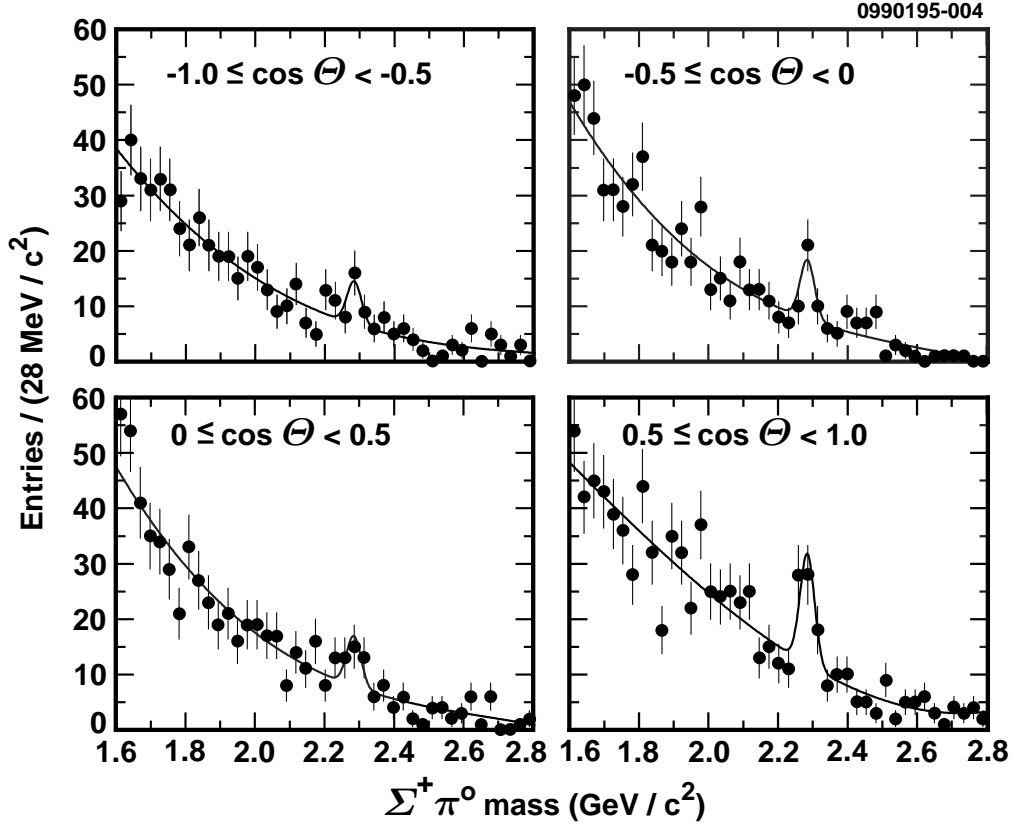,width=5.5in}
  }
  \caption{ $\Sigma^+\pi^0 $ Invariant mass distribution in four $\cos\Theta$
bins}
  \label{fig:sigp0cos}
\end{figure}

As in the $\Lambda\pi^+ $ case, we divide the data into four \COST\
bins.  The four different invariant mass distributions can be seen in
Figure~\ref{fig:sigp0cos}. An increase in the number of \SIGPI\ events
occurs as \COST\ increases.

\begin{figure}
 \centerline{
 \psfig{bbllx=90pt,bblly=390pt,bburx=490pt,bbury=720pt,file=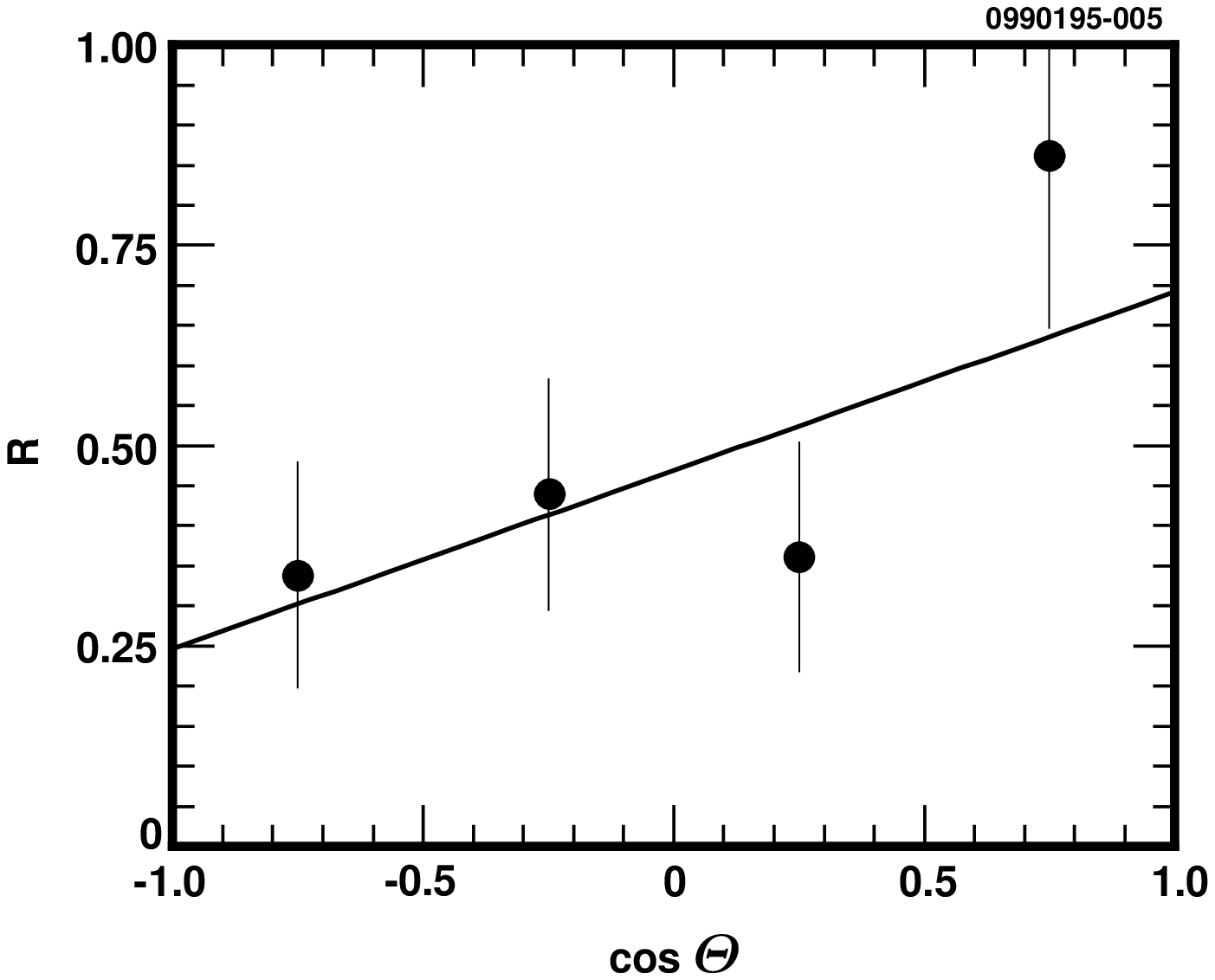}
 }
  \caption{ Efficiency corrected yields vs $\cos\Theta$ for decay \LCsigpi}
  \label{fig:sigp0asy_show}
\end{figure}

The normalized, efficiency-corrected $\Sigma^+\pi^0 $ yield as a function of
\COST\ is plotted in Figure~\ref{fig:sigp0asy_show}. We fit the plot to a
straight line and find a slope of $0.22 \pm 0.15$ with a confidence level of
$28\%$.  The $\Sigma^+ $ asymmetry is large, $\alpha_{\Sigma^+} = -0.98 \pm
0.015$.  We therefore obtain $\alpha_{\Lambda_c} = -0.45 \pm 0.31$, where the
error is statistical only.

We have considered systematic errors similar to those in the \LAMPI \ case.
The efficiency for the $\pi^0 $ coming from the
$\Sigma^+ $ changes very slowly with momentum, since the electromagnetic
calorimeter is efficient for photons above 30 MeV/c. Hence the efficiency is
flat as a function of \COST .  The fitting of the \SIGPI \ invariant mass
distributions is less sensitive to the background shape since there are no
reflections from other particles.  The total systematic error on $\aLC $ is
$\pm 0.06$ which is significantly smaller than in the \LAMPI \ case.

Based on factorization, Bjorken\cite{BJ} first predicted $\aLC $ to be
$-1$ for \LClampi.  Since then, the two-body baryon-pseudoscalar
decays of charmed baryons have been studied by a number of authors. Theoretical
models predict the parity-conserving and parity-violating
amplitudes, and from these the asymmetry and the decay rate can be
calculated. Therefore, a useful model should correctly predict both
the asymmetry and the decay rate. Since both $B(\Lambda\pi^+)/B(pK^-\pi^+)
$\cite{OLDLAMPI}
and $B(\Sigma^+\pi^0)/B(pK^-\pi^+) $\cite{Procario} have been measured, we can
make simultaneous comparison, which is shown in table~\ref{tab:comptheory}.
The predictions of  \.Zenczykowski agree best with the complete set
of data. They are within $1\sigma $ for all quantities except the decay
width of $\Sigma^+\pi^0$, which is within $2\sigma $. Three of the five
models have a strong disagreement with the $\Sigma^+\pi^0 $ asymmetry,
while are in better agreement with the other quantities. The predictions
of Uppal, Verma and Khanna are quite good for asymmetries but poor for
decay widths.

If a model fails to predict the decay width and asymmetry correctly, it can
be instructive to compare its predictions of the $s-$wave and $p-$wave
amplitudes, \A and \B, with those found from the experiment.  We ignore
any final-state interactions as do the theoretical models.
Since \A and \B always appear quadratically we will get two sets
of solutions. It is necessary to measure the angle between the initial and
final baryon spins to discriminate between these solutions, which we are
unable to do. The results are shown in Table~\ref{tab:AB}.

Cheng and Tseng's calculation agrees well on 3 of the 4 amplitudes in the
first set. It only fails on \A in the $\Sigma^+\pi^0 $ case, where it has the
wrong sign.  Both Cheng and Tseng, and Xu and Kamal claim that \A in
$\Sigma^+\pi^0 $ is sensitive to the on-shell correction.

In conclusion, we have made the first measurement of the decay asymmetry
in \LCsigpi and an improved measurement of the decay asymmetry
in  \LClampi.  Most model predictions agree with the \LClampi
measurements, while the \LCsigpi asymmetry measurement seems to be the
hardest to predict.

We gratefully acknowledge the effort of the CESR staff in providing
us with excellent luminosity and running conditions.  This work was
supported by the National Science Foundation, the U.S. Dept. of
Energy, the Heisenberg Foundation, the SSC Fellowship Program of
TNRLC, and the A.P.  Sloan Foundation.

\begin{table}
\begin{center}
\begin{tabular}{p{2.0in}cccc}
                                  & \multicolumn{2}{c}{\LClampi} &
\multicolumn{2}{c}{\LCsigpi}  \\
                                  & $\Gamma (10^{11} {\rm s}^{-1})$  &  $\alpha
$ & $ \Gamma (10^{11} {\rm s}^{-1})$ & $\alpha $  \\ \hline
CLEO                              & $ 0.40\pm 0.11 $
                                  & $-0.94\ ^{+\ 0.21\ +\ 0.12}_{-\ 0.06\ -\
0.06}$ 
                                  & $ 0.44\pm 0.12 $
                                  & $-0.45 \pm 0.31 \pm 0.06 $  \\ \hline
Xu \& Kamal \cite{XK}	          & $0.81$     & $-0.67$    &  $ 0.17$   &
$0.91$ \\ \hline
Cheng \& Tseng \cite{CT}          & $0.46$     & $-0.96$    &  $ 0.38$   &
$0.83$ \\ \hline
K\"{o}rner \& Kr\"{a}mer\cite{KK} & $0.37$     & $-0.70$    &  $ 0.16$   &
$0.71$ \\ \hline
Uppal, Verma \& Khanna\cite{UVK}  & $1.17$     & $-0.85$    &  $ 1.22$   &
$-0.32$ \\ \hline
\.Zenczykowski\cite{Zen}          & $0.31$     & $-0.86$    &  $ 0.23$   &
$-0.76$ \end{tabular}
\end{center}
 \caption{ Comparison of experimental decay widths and asymmetries with
model predictions. Experimental widths are derived from CLEO relative
branching ratios and PDG 94 values of the $\Lambda_c^+ $ lifetime and
$pK^-\pi^+ $ branching fraction. Theoretical predictions that were published
as branching fractions have been converted to decay widths using $\Lambda_c^+ $
lifetime.}
\label{tab:comptheory}
\end{table}

\begin{table}
 \begin{center}
  \begin{tabular}{lcccc}
     & \multicolumn{2}{c}{\LClampi} & \multicolumn{2}{c}{\LCsigpi} \\ \hline
        &  A            &         B        &       A        &    B  \\ \hline
CLEO II & $-3.0\ ^{+0.8}_{-1.2} $ & $ 12.7\ ^{+2.7}_{-2.5}$ & $ 1.3\
^{+0.9}_{-1.1} $ & $-17.3\ ^{+2.3}_{-2.9}$ \\
        & $-4.3\ ^{+0.8}_{-0.9} $ & $  8.9\ ^{+3.4}_{-2.4}$ & $ 5.4\
^{+0.9}_{-0.7} $ & $ -4.1\ ^{+3.4}_{-3.0}$ \\
Xu \& Kamal \cite{XK}              & $-2.7$ & $20.8$ &  $-2.9$ & $  -6.0$   \\
Cheng \& Tseng\cite{CT}            & $-3.5$ & $13.2$ &  $-2.4$ & $ -14.6$   \\
K\"orner \& Kr\"amer\cite{private} & $-1.9$ & $13.9$ &  $-1.3$ & $  -9.9$   \\
\end{tabular}
\end{center}
\caption{Comparison of the measurements of the $s$ and $p$ wave amplitudes,
A and B, in \LClampi and \LCsigpi with theoretical predictions. The
amplitudes have all been converted to common units of
$G_FV_{cs}V_{ud}\times 10^{-2} $ GeV$^2$. }
\label{tab:AB}
\end{table}


\begin{references}
\bibitem{Charge} Throughout this paper the charge conjugate
reaction is implied.
\end{references}
\end{document}